\documentclass{article}
\usepackage{graphicx}

\usepackage[utf8]{inputenc}
\usepackage{amsmath, amssymb, mathtools, graphicx, amsthm, dsfont, float, bbm, hyperref, caption, adjustbox, kantlipsum, algorithm, authblk, algcompatible, fullpage, subcaption}

\usepackage{xcolor}
\usepackage{comment}
\usepackage{cleveref}
\usepackage{booktabs}
\usepackage[round]{natbib}
\usepackage{paralist}

\newtheorem{theorem}{Theorem}
\newtheorem{proposition}{Proposition}

\theoremstyle{definition}
\newtheorem{definition}{Definition}
\newtheorem{assumption}{Assumption}

\newcommand{\Cov}{\mathrm{Cov}}
\newcommand{\bbN}{\mathbb{N}}
\newcommand{\bbR}{\mathbb{R}}
\newcommand{\bbE}{\mathbb{E}}
\newcommand{\bbP}{\mathbb{P}}
\newcommand{\Var}{\mathrm{Var}}
\newcommand{\st}{\text{ } \vert \text{ } }

\newcommand{\simind}{\stackrel{\mathrm{ind}}{\sim}}

\newcommand{\red}{\mathrm{red}}
\newcommand{\IPW}{\mathrm{IPW}}
\newcommand{\bern}{\mathrm{Bern}}

\title{Reservoir Designs for Online Paired Experiments}
\author{Tim Morrison}
\date{May 2025}

\begin{document}

\maketitle

\begin{abstract}
We study the question of how best to stratify units into matched pairs in online experiments, so that units within a pair receive opposite treatment. Past work by \cite{bai-romano2022} has demonstrated the asymptotic variance improvement that comes from pairing units with similar covariates in this way. However, their method requires knowing the covariates for all units a priori; this is not the case in many A/B testing problems, in which units arrive one at a time and must have treatment assigned immediately. Inspired by the terminology of \cite{KK14}, we thus introduce the notion of a \textit{reservoir design}, which maintains a reservoir of unpaired units that can potentially be paired with an incoming unit. We construct a particular reservoir design that uses a distance-based criterion to determine pairing and, via a packing argument, prove conditions under which it attains the asymptotic variance improvement of \cite{bai-romano2022}. We illustrate our reservoir design on synthetic and semi-synthetic examples and find improved performance relative to both IID sampling and the design of \cite{KK14}.

\end{abstract}

\section{Introduction} \label{sec:intro}
Consider an online experiment in which units with covariates $X_t \in \bbR^d$ arrive one at a time, have a binary treatment $Z_t \in \{0, 1\}$ assigned immediately, and have a response of interest $Y_t \in \bbR$ measured some time later. The aim is then to estimate the average effect of the treatment in the population. This is the standard procedure for A/B testing, which has become ubiquitous across many online platforms \citep{kohavi2020}. In A/B tests, practitioners often advise that units be independently assigned to the treatment group with probability $50\%$ to improve interpretability, robustness, or statistical power \citep{kohavi2015, georgiev2021, li2022, doordash2023}. However, the sequential nature of the experiment and the presence of covariates suggests that there may be simple ways of improving on the IID 50/50 assignment. 

In this paper, we consider matched-pair designs, in which units with similar covariates are grouped into pairs and exactly one of them is given treatment. This maintains the same marginal treatment probability of $50\%$ for each unit while introducing weak treatment correlations to increase covariate balance between the treatment and control groups. Support for matched-pair designs comes from \cite{bai2022}, who proved that, among all stratified designs that retain the same $50\%$ marginal treatment probability, a particular matched-pair design minimizes the variance of the inverse propensity weighting (IPW) estimator of the treatment effect. This optimal design depends on oracle estimation and is thus not feasible to implement, but \cite{bai-romano2022} showed that certain matched-pair designs that are constructible from the covariates achieve the same benefit asymptotically. However, these designs still require knowing all covariates a priori, which is not the case in online experiments. 

In order to pair units in an online experiment, there must be some units that are unpaired on their arrival. This motivates the idea of a \textit{reservoir design}, in which the experimenter maintains a reservoir of unpaired units that are candidates for pairing with future arrivals. Intuitively, a new unit should be paired with a candidate in the reservoir if they have similar covariates. This introduces a tradeoff in the stringency of the pairing criterion: a more lenient criterion will result in more (but worse) pairings, and a more stringent one will result in a fewer (but better) pairings. 

Using an argument based on packing numbers, we outline a particular pairing criterion that achieves the same asymptotic benefit as the designs of \cite{bai-romano2022}. This criterion computes the Euclidean distances between a new unit's covariates and those of each unit in the reservoir. It then pairs the new unit with its nearest neighbor in the reservoir if and only if the distance between them falls below some cutoff radius. This radius decays to zero with sample size, ensuring that the matchings improve over time. Conveniently, our method does not require knowing the final sample size at the beginning of the experiment, which need not be the case; for example, an A/B test may run for a particular duration rather than for a particular number of units. 

To our knowledge, the only line of past work that has studied pairing in online experiments is that of \cite{KK14}. They propose using a pairing criterion based on the Mahalanobis distance to the nearest unit in the reservoir. While their method is sensible and effective, they study it only under a strong model in which there is no treatment effect heterogeneity and the errors are Gaussian and homoscedastic. In the general case, their method does not attain the asymptotic result of \cite{bai-romano2022}. Moreover, the cutoff radius for the Mahalanobis distance is a free parameter in their method, and there is no obvious choice for how it should be set. In a follow-up paper \citep{KK23}, the authors also construct a variant of their method that replaces the Mahalanobis distance with a weighted sum of coordinate-wise distances, where the weights are iteratively updated via a particular regression of $Y$ on $(X, Z)$. 

While pairing in online experiments has been minimally explored, many alternative forms of covariate balance have been adapted to the sequential setting. \cite{efron1971} proposes biased-coin designs that update the treatment probability depending on the current proportion of treated units, which has been generalized in several papers (e.g., \cite{atkinson1982, smith1984, antognini2012}) to allow for covariate adaptivity. \cite{pocock1975} and \cite{hu2012} both present sequential randomization methods when covariates are categorical that aim to balance treatment allocation within all covariate strata. In the more general case, \cite{zhao2025} proposes ``pigeonhole designs" that grid the covariate space into hypercubes and balance the number of treated and control units in each one. Several papers, beginning with \cite{zhou2018} and including \cite{zhu2023}, adapt the rerandomization ideas of \cite{morgan2012} to the sequential setting to ensure covariate balance. When new units arrive in pairs, the pairwise sequential randomization procedure of \cite{qin2016} assigns opposite treatment to them, with probabilities dependent on the new Mahalanobis distance between the treatment and control groups. 

This paper is organized as follows. In Section \ref{sec:setup}, we introduce the key notation and formally define our notion of a reservoir design. In Section \ref{sec:methods}, we provide conditions under which a reservoir design achieves the central limit theorem of \cite{bai-romano2022} and outline a particular design that does so. In Section \ref{sec:sims}, we present several simulations to assess the performance of our design in comparison to IID sampling and the ``on-the-fly" design of \cite{KK14}. These include a semi-synthetic example based on a real experimentation dataset from the online advertising company Criteo. We include all proofs in Appendix \ref{sec:proofs}. 

\section{Setup} \label{sec:setup}
We assume that covariates $X_t \in \bbR^d$ arrive sequentially and IID from some distribution $P_X$. Upon arrival of $X_t$, we must immediately assign a binary treatment $Z_t \in \{0, 1\}$. We do so in a way such that $\bbP(Z_t = 1 \st X_t) = 1/2$, which we write as $\bern(1/2)$. Some time later, we observe a response $Y_t \in \bbR$. This process terminates after $T$ units, where $T$ may be known or unknown a priori. After all responses have been observed, we estimate the average treatment effect $\tau = \bbE[Y(1) - Y(0)]$ via the inverse propensity weighting (IPW) estimator 
\begin{equation} \label{eq:IPW-def} \hat{\tau}_{\IPW}(T) = \frac{2}{T} \sum_{t = 1}^{T} Z_t Y_t - (1 - Z_t)Y_t. \end{equation} 
Note that $\hat{\tau}_{\IPW}(T)$ is unbiased for $\tau$ regardless of the particular joint distribution of $(Z_1, \ldots, Z_T)$, so long as each $Z_t$ is marginally $\bern(1/2)$. 

We write $\mu_1(X_i) = \bbE[Y_i(1) \st X_i]$, and analogously for $\mu_0(X_i)$. Similarly, we write $\sigma_1^2(X_i) = \Var(Y_i(1) \st X_i)$, and analogously for $\sigma_0^2(X_i)$. As in \cite{bai2022}, we write $g(X_i) = \bbE[Y_i(1) + Y_i(0) \st X_i]$ for the sum of conditional outcome means given covariates. 

We consider designs that pair indices and assign opposite treatments to those units. Intuitively, it is best to pair units with similar covariates $X$, which reduces variance due to covariate imbalance. However, in this problem we do not have access to all covariates at once. Upon arrival of unit $t$, we must choose whether to pair it with an unpaired unit that we have already seen or assign it a new IID treatment. 

This line of reasoning leads to the notion of a \textit{reservoir design}, in which we maintain a reservoir of units that have yet to be paired. When a new unit arrives, we can either assign it an IID $\bern(1/2)$ treatment status and add it to the reservoir or pair it with a unit $Z_s$ already in the reservoir and assign $Z_t = 1 - Z_s$. In the latter case, $Z_s$ would then be removed from the reservoir. We formalize this in the definition below. 

\begin{definition}\label{def:reservoir-design}
    A \textit{reservoir design} is any distribution on $(Z_1, Z_2, \ldots, Z_T) \in \{0, 1\}^T$ such that: 
    \item (1)  $Z_1 \sim \bern(1/2)$,
    \item (2) at step $t$, $Z_t$ is either 
        \item \quad (a) assigned as an IID $\bern(1/2)$ value and added to the reservoir, or 
        \item \quad (b) given the opposite treatment to a unit in the reservoir, which is then removed from the reservoir, and
    \item (3) the choice in step (2) is a deterministic function of $(X_1, X_2, \ldots, X_t)$. 
\end{definition}

Condition (3) guarantees that, conditional on all covariates $X \in \bbR^{T \times d}$, there is no remaining randomness in the pairings. We borrow the term ``reservoir" from \cite{KK14}, whose ``on-the-fly" design is a particular example of what we call a reservoir design. We also choose this name because of the similarity to reservoir sampling (e.g., \cite{vitter1985}), in which one selects $n$ units uniformly from a sequential stream of $N$ of them (with $N$ unknown) by maintaining a reservoir that is randomly updated. 

After $Z_t$ has been assigned, we write $\mathcal{R}_t$ for the set of indices in the reservoir and $\mathcal{M}_t$ for the set of pairs of matched indices. We write $\mathcal{F}_t = \sigma(X_1, X_2, \ldots, X_t)$ for the $\sigma$-algebra generated by the covariates up to and including unit $t$; by Definition \ref{def:reservoir-design}, this is sufficient to determine $\mathcal{R}_t$ and $\mathcal{M}_t$. 

Conveniently, the general formula for $\Var(\hat{\tau}_{\IPW}(T) \st X)$ is the same for any reservoir design, as we show in the following proposition. 

\begin{proposition} \label{prop:variance-formula}
If treatment $Z$ is assigned via a reservoir design, then 
\begin{equation} \label{eq:variance_formula}
\Var(\hat{\tau}_{\IPW}(T) \st X) = \frac{1}{T^2} \sum_{i = 1}^{T} \left(2\sigma_1^2(X_i) + 2\sigma_0^2(X_i) + g(X_i)^2\right) - \frac{2}{T^2} \sum_{(i, j) \in \mathcal{M}_T} g(X_i)g(X_j).
\end{equation}
\end{proposition}
The proofs of Proposition \ref{prop:variance-formula} and all subsequent results are in Appendix \ref{sec:proofs}. Proposition \ref{prop:variance-formula}, which resembles and generalizes some results in \cite{bai2022},  clarifies how pairing can reduce variance. The first term in \ref{eq:variance_formula} is the variance that would arise from assigning each $Z_t \simind \bern(1/2)$, or equivalently, from putting every unit in the reservoir and making no pairings. The variance improvement from pairing is given by the term subtracted in \ref{eq:variance_formula}, which is positive if $g(X_i)$ and $g(X_j)$ tend to have the same sign and large if they tend to be similar in magnitude. Intuitively, this should be the case if $g(X)$ is continuous and units with nearby covariates are paired. If the reservoir is empty, then 
\begin{equation} \label{eq:var-empty-res} \Var(\hat{\tau}_{\IPW}(T) = \frac{1}{T^2} \sum_{(i, j) \in \mathcal{M}_T} 2\sigma_1^2(X_i) + 2\sigma_0^2(X_i) + (g(X_i) - g(X_j))^2,\end{equation}
in which case the design is invariant to additive shifts in the $g(X)$ function. 

Proposition \ref{prop:variance-formula} also highlights the difficult trade-off in this problem between quality and quantity of pairings. If we make more pairings, then there will be more terms in the second sum of \ref{eq:variance_formula}, but the average variance reduction from each pairing may be worse. In the next section, we adapt and leverage some results from \cite{bai-romano2022} to try to optimize this trade-off and improve on the reservoir design of \cite{KK14}. 

\section{Constructing reservoir designs} \label{sec:methods}
In this section, we develop an algorithm for pairing units with asymptotic guarantees in a sequential setting. In Section \ref{ssec:general-results}, we first state general assumptions under which a reservoir design recovers the central limit theorem of \cite{bai-jiang2024}. In Section \ref{ssec:packing-design}, we then present a simple algorithm that satisfies these assumptions and, up to some mild conditions, is agnostic to the distribution $P_X$ and the sample size $T$. 

\subsection{General results} \label{ssec:general-results}

We begin with some assumptions on the joint distribution of $(X_i, Y_i(0), Y_i(1))$, which are similar to those in \cite{bai-romano2022}. 
\begin{assumption} \label{assumptions:XY}
    \item (a) $\bbE[||X_i||_2^4] < \infty$. 
    \item (b) $\bbE[\Var(Y(z) \st X)] > 0$ for $z \in \{0, 1\}$. 
    \item (c) $\bbE[Y_i^r(z) \st X = x]$ is Lipschitz in $x$ for $z \in \{0, 1\}$ and $r \in \{1, 2\}$. 
    \item (d) The distributions of $Y_i(1)$ and $Y_i(0)$ are both sub-Gaussian.
\end{assumption}
 The first assumption is a standard moment condition. The second assumption states that there is still some variability left in the potential outcomes after conditioning on $X$. The third assumption ensures that nearby $X$ values have similar conditional mean outcomes, which justifies pairing similar units for variance reduction. The fourth assumption is not included in \cite{bai-romano2022}, but we need it to ensure that the outcomes are not adversarially larger in the reservoir than in the matched pairs.

Let $n_{\mathcal{R}}(t) = |\mathcal{R}_t|$ and $n_{\mathcal{M}}(t) = |\mathcal{M}_t|$ be the number of units in the reservoir and in matched pairs, respectively, after $Z_t$ has been assigned. By decomposing the estimator $\hat{\tau}_{\IPW}(T)$ into the terms in the reservoir and in the matched pairs, we can write it as 
\begin{align*} 
\hat{\tau}_{\IPW}(T) &= \frac{2}{T} \sum_{i = 1}^{T} (2Z_i - 1)Y_i \\ 
&=  \frac{2}{T} \sum_{(i, j) \in \mathcal{M}_T} \left((2Z_i - 1)Y_i + (2Z_j - 1)Y_j\right) + \frac{2}{T} \sum_{i \in \mathcal{R}_T} (2Z_i - 1)Y_i \\ 
&=  \frac{n_{\mathcal{M}}(T)}{T} \hat{\tau}_{\mathcal{M}}(T) + \frac{n_{\mathcal{R}}(T)}{T} \hat{\tau}_{\mathcal{R}}(T) \\ 
&= \hat{\tau}_{\mathcal{M}}(T) + \frac{n_{\mathcal{R}}(T)}{T}\left(\hat{\tau}_{\mathcal{R}}(T) - \hat{\tau}_{\mathcal{M}}(T)\right).
\end{align*}
where $\hat{\tau}_{\mathcal{R}}(T)$ is the IPW estimator averaged over only the units in the reservoir, and analogously for $\hat{\tau}_{\mathcal{M}}(T)$. In the last line, we use that $n_{\mathcal{M}}(T) = T - n_{\mathcal{R}}(T)$.

\cite{bai-romano2022} study matched-pair designs in the case where every $X_i$ is known a priori, so there is no need for a reservoir and each unit is paired. Under Assumption \ref{assumptions:XY}, they prove a CLT for this matching procedure that yields a variance reduction over independent assignment. If our procedure ended with a reservoir of size zero, we could simply apply their theorem to get the same result. However, because 
$$\sqrt{T}(\hat{\tau}_{\IPW}(T) - \tau) =  \sqrt{T} \left(\hat{\tau}_{\mathcal{M}}(T) - \tau \right) + \frac{n_{\mathcal{R}}(T)}{\sqrt{T}}\left(\hat{\tau}_{\mathcal{R}}(T) - \hat{\tau}_{\mathcal{M}}(T) \right),$$
it suffices merely that $n_{\mathcal{R}}(T)$ does not grow too quickly. In addition, the other necessary condition is that the matches improve in quality, meaning that the average intra-pair distance between covariates tends to zero. We combine this all into the following theorem.

\begin{theorem} \label{thm:CLT}
Suppose that Assumption \ref{assumptions:XY} holds, that $n_{\mathcal{R}}(T) = o_p(\sqrt{T/\log(T)})$, and that 
\begin{equation} \label{eq:intra-distances} \frac{1}{T} \sum_{(i, j) \in \mathcal{M}_T} ||X_i - X_j||_2^r \overset{p}{\to} 0 \end{equation}
as $T \to \infty$ for both $r \in \{1, 2\}$. Then 
$$\sqrt{T}(\hat{\tau}_{\IPW} - \tau) \overset{d}{\to} N(0, \sigma_{\red}^2),$$
where
\begin{equation} \label{eq:sigmasq-red} \sigma_{\red}^2 = \sigma_1^2 + \sigma_0^2 - \frac{1}{2} \Var(g(X)).\end{equation}
\end{theorem}
Theorem \ref{thm:CLT} recovers the central limit theorem of \cite{bai-romano2022} provided that the reservoir grows at a rate slightly slower than $\sqrt{T}$ and that the average intra-pair covariate distances shrink to zero. We present a specific design that meets these criteria in the next subsection. Of note, the design of \cite{KK14} does not satisfy the intra-pair distances condition \ref{eq:intra-distances}, as discussed further in the next section. The limiting variance $\sigma^2_{\red}$ in \ref{eq:sigmasq-red} has a simple interpretation: the variance improvement over an IID design is larger if $g(X)$ is more variable, since pairing accounts for a greater proportion of the variation in $Y$. 

The proof of Theorem \ref{thm:CLT} essentially involves showing that the reservoir does not matter asymptotically and then appealing to Lemma S.1.4 of \cite{bai-romano2022}. We note also that their consistent estimator for $\sigma^2_{\red}$ works just as well in our case, so we could apply their Theorem 3.3 to obtain an asymptotically valid confidence interval for $\tau$.

\subsection{Attaining the CLT via packing} \label{ssec:packing-design}
In this section, we construct an explicit reservoir design for which Theorem \ref{thm:CLT} holds. Our procedure shares similarities with that of \cite{KK14}, but we use a packing number argument to attain the desired asymptotic behavior. 

Suppose first that $T$ is known and $P_X$ is supported on $[0, 1]^d$. Suppose that data for a new index $t$ has arrived. If the reservoir is empty, we must assign $Z_t \simind \bern(1/2)$. Otherwise, let $\text{nn}(t)$ be the index of the nearest neighbor in the reservoir, measured in Euclidean distance. Consider the procedure that pairs $t$ with $\text{nn}(t)$ if and only if $||X_t - X_{\text{nn}(t)}||_2 < \lambda_T$. This is very similar to the procedure of \cite{KK14}, except they use Mahalanobis distance and take the cutoff to depend on $t$, not $T$. 

We now show how to choose $\lambda_T$ to obtain the desired asymptotic behavior. The key observation is that, after all $T$ units have arrived and the matching is complete, the points remaining in the reservoir will form a $\lambda_T$-packing of $[0, 1]^d$. That is, no two points in the reservoir can be within $\lambda_T$ of each other in Euclidean distance, or else they would have been paired when the later one arrived. Hence, the size of the reservoir can be at most $O(\lambda_T^{-d})$ by a standard result about the packing number of $[0, 1]^d$ (see, e.g., Lemma 2.5 of \cite{vandegeer}). Choosing $\lambda_T$ so that the size of the reservoir is $o_p\left(\sqrt{T/\log(T)}\right)$, in accordance with Theorem \ref{thm:CLT}, we obtain
$$\lambda_T^{-d} < \log(T)^{-1/2} T^{1/2} \iff \lambda_T > \log(T)^{1/2d} T^{-1/2d}.$$
So long as this condition holds and $\lambda_T \to \infty$ as $T$ grows, the conditions of Theorem \ref{thm:CLT} are satisfied. For instance, one could choose $\lambda_T = T^{-1/(2 + \delta)d}$ for some small $\delta > 0$. To avoid the units of $X$ mattering, we also propose standardizing the columns of $X$ to have sample mean zero and sample standard deviation one before comparing distances. 

This method can be easily adapted to the case in which $T$ is not known a priori. At index $t$, simply run the above method with a pairing radius of $t^{-1/(2 + \delta)d}$. When the final index $T$ is reached, cease the procedure. Intuitively, the reservoir size from this procedure can be no larger than the reservoir size from the procedure run at the smaller cutoff radius where $t = T$, so the total reservoir size should be $o_p(\sqrt{T/\log(T)})$. We make this precise in the proof of Theorem \ref{thm:CLT-packing}.

\begin{algorithm}[t] 
\caption{Reservoir design via packing argument} \label{alg:packing-design}
\begin{algorithmic}[1]
\STATE{ \textbf{Input:} exponent offset $\delta > 0$, covariate dimension $d \in \bbN$} 
\STATE{\textbf{Set} $\mathcal{R}_0 = \{\}$ and $t = 1$}
\WHILE{experiment running}
    \STATE Collect $X_t \in \bbR^d$ and append to $X$
    \STATE Normalize $X \in \bbR^{t \times d}$ to have column means zero and variance one
    \STATE Set $\lambda_t = t^{-1/(2 + \delta)d}$
    \IF{$\mathcal{R}_{t - 1}$ = \{\}}
        \STATE $Z_t \simind \bern(1/2)$ 
        \STATE $\mathcal{R}_t = \{t\}$
    \ELSE
        \STATE $\text{nn}(t) = \underset{s \in \mathcal{R}_{t - 1}}{\text{argmin }} ||X_t - X_s||_2$
        \IF{$||X_t - X_{\text{nn}(t)}||_2 < \lambda_t$}
            \STATE $Z_t = 1 - Z_{\text{nn}(t)}$ 
            \STATE $\mathcal{R}_t = \mathcal{R}_{t - 1}\setminus{\{\text{nn}(t)\}}$
        \ELSE
            \STATE  $Z_t \simind \bern(1/2)$ 
            \STATE $\mathcal{R}_t = \mathcal{R}_{t - 1} \cup \{t\}$
        \ENDIF
    \ENDIF
    \IF{experiment done} 
        \STATE break
    \ENDIF
    \STATE{$t = t + 1$}
\ENDWHILE
\STATE{\textbf{Collect} response data $Y_t \in \bbR$ for $t \leq T$}
\STATE{\textbf{Compute}}
$$\hat{\tau}_{\IPW} = \frac{2}{T} \sum_{t = 1}^{T} (2Z_t - 1)Y_t$$
\STATE{\textbf{Output:} $\hat{\tau}_{\IPW}$}
\end{algorithmic}    
\end{algorithm}

We summarize this method in Algorithm \ref{alg:packing-design}. Theorem \ref{thm:CLT-packing} below rigorously proves many of the details that we have argued heuristically here. In addition, it generalizes beyond the case in which $P_X$ has compact support to the case in which $X$ is sub-Gaussian.   

\begin{theorem} \label{thm:CLT-packing}
    Suppose that Assumption \ref{assumptions:XY} holds and that the distribution of $X$ is sub-Gaussian. Then the procedure in Algorithm \ref{alg:packing-design} satisfies the conditions of Theorem \ref{thm:CLT}, and so 
    $$\sqrt{T}(\hat{\tau}_{\IPW} - \tau) \overset{d}{\to} N(0, \sigma^2_{\red}),$$
    where $\sigma^2_{\red}$ is as in \ref{eq:sigmasq-red}. 
\end{theorem}

In contrast to Algorithm \ref{alg:packing-design}, the reservoir design of \cite{KK14} (henceforth, KK14) computes the Mahalanobis distances $(X_t - X_j)^{\top} \hat{\Sigma}_{t-1}^{-1}(X_t - X_j)$ between a new unit $X_t$ and all other units in the reservoir, where $\hat{\Sigma}_{t-1}$ is an estimate of the covariance matrix from the first $t - 1$ units. To handle difficulties in estimating $\Sigma$ in finite samples, they also use a burn-in phase of some size $n_0$ for which all initial units are added to the reservoir. Because $n_0$ does not matter asymptotically, we take $n_0 = d$ for simplicity when running their method and use the sample covariance matrix for all subsequent iterations. Nonetheless, we note that $n_0$ is an important hyperparameter for KK14 and, in their paper, is often chosen to be a good deal larger. 

If the covariates are Gaussian, then for $t > d$
$$(X_t - X_j)^{\top} \hat{\Sigma}_{t-1}^{-1}(X_t - X_j) \sim \frac{2d(t - 1)}{t - d} F_{d, t - d}.$$
Hence, they use a cutoff radius (in Mahalanobis distance) of $\frac{2d(t-1)}{t - d} F_{\lambda, d, t - d}$, where $\lambda \in (0, 1)$ is a fixed constant and $F_{\lambda, d, t - d}$ is the $\lambda$ quantile of the $F_{d, t - d}$ distribution. While this limit does vary with $t$, it asymptotes to a nonzero constant, since 
$$\underset{t \to \infty}{\lim} \text{ } \frac{2d(t-1)}{t - d} F_{\lambda, d, t - d} = 2\chi^2_{\lambda, d}.$$
This equation uses that an $F_{d, n - d}$ random variable converges in distribution to a $\chi^2_d/d$ random variable as $n$ tends to infinity. Hence, the average intra-pair distance for their reservoir design does not converge to zero, so they do not attain the CLT of \cite{bai-romano2022}. 

\begin{figure}[t]
    \begin{subfigure}[b]{0.5\textwidth}
        \centering
        \includegraphics[width=\textwidth]{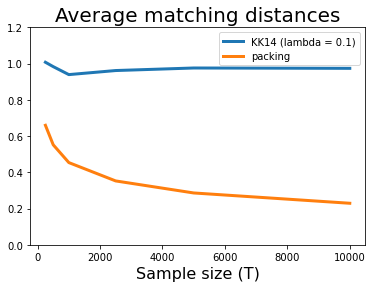}
    \end{subfigure}
    \hfill
    \begin{subfigure}[b]{0.5\textwidth}
        \centering
        \includegraphics[width=\textwidth]{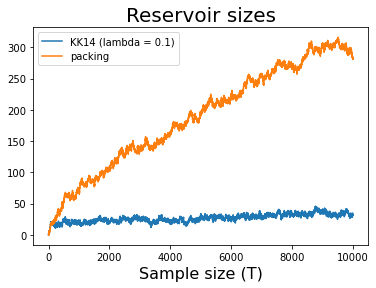}
    \end{subfigure}
\caption{Average intra-pair distances among matched pairs (left), and current reservoir size (right), for the \cite{KK14} method (KK14) and our packing method. Covariates are correlated Gaussians on $\bbR^3$.}
\label{fig:dist-res-comps}
\end{figure}

We can also see this difference in Figure \ref{fig:dist-res-comps}, which is based on a simulation with covariates that are correlated Gaussians on $\bbR^3$. Specifically, we take $X_t \simind N(\mu, \Sigma)$, with 
$$\mu = \begin{bmatrix} 1 \\ 2 \\ 3 \end{bmatrix} \text{and } \Sigma = \begin{bmatrix} 1 & 0.5 & 0.2 \\ 0.5 & 2 & 0.3 \\ 0.2 & 0.3 & 1.5 \end{bmatrix}.$$ 
The left plot compares the average intra-pair distances (among paired units) between the KK14 method and our pairing method, and the right plot compares the size of the reservoir at time $t$. As we can see, the average intra-pair distance for their method is essentially flat, whereas ours steadily decreases and is much lower. In addition, their reservoir size does not grow with $T$, whereas ours grows at a sufficient rate to ensure proper asymptotics. 

The cutoff radius of (roughly) $t^{-1/2d}$ implies a curse of dimensionality in finite samples. However, we emphasize that not every covariate needs to be used; the CLT of Theorem \ref{thm:CLT-packing} holds for whatever covariates the experimenter chooses to use, which may only be the ones that they expect to be most predictive. In practice, one could also iteratively update the algorithm to match on the first few principal components of the covariate data seen so far. In addition, if $Y$ data is available, one could use Lasso or another feature selection algorithm and match on just the selected features. 

\section{Simulation results} \label{sec:sims}
In this section, we present several simulations to compare the performance of our packing-based method and the KK14 method. In each simulation, we vary the sample size and consider five different designs: \\ \\ 
(1) IID treatment assignment \\ 
(2) KK14 with $\lambda = 0.1$ \\ 
(3) Our packing method \\ 
(4) Packing based on the (unknown) true $g(X)$ \\\ 
(5) The optimal matched-pairs design of $\cite{bai2022}$. \\ \\
Of these, the first three can be implemented using the known covariate data, whereas the last two require oracle knowledge of the true $g(X)$ function. Method (4) pairs two units if their $g(X)$ terms are sufficiently close, whereas method (5) sorts units by $g(X)$ and pairs consecutive units, resulting in a reservoir of size zero. This method also requires knowing all the data a priori, whereas the first four can all be done sequentially. For this reason, it is essentially a lower bound on the performance of a pairing-based design.

We take $\delta = 0$ in Algorithm \ref{alg:packing-design} for simplicity, meaning a cutoff radius at step $t$ of $t^{-1/2d}$. While this technically does not provide the asymptotics of Theorem \ref{thm:CLT-packing}, there would be virtually no difference in finite sample if we chose a vanishingly small $\delta > 0$. We choose $\lambda = 0.1$ for the KK14 method because it is what they choose for their own simulations, and because we found it to be a strong choice for their free parameter.  \\ \\
We also consider a few simulation settings: \\ \\ 
(1) $X_t \simind \text{Unif}[0, 1]^2$, and $Y_t = Z_t (\textbf{1}_2^{\top} X_t)^2 + N(0, 0.01)$. \\ 
(2) $X_t \simind N(\mu, \Sigma)$, with 
$$\mu = \begin{bmatrix} 1 \\ 2 \\ 3 \end{bmatrix} \text{and } \Sigma = \begin{bmatrix} 1 & 0.5 & 0.2 \\ 0.5 & 2 & 0.3 \\ 0.2 & 0.3 & 1.5 \end{bmatrix},$$ 
and $Y_t = \sin(\textbf{1}_3^{\top}X_t) + Z_t(3 + 2 \cos(\textbf{1}_3^{\top}X_t)) + N(0, 0.01)$. \\ 

Here, $\textbf{1}_d$ is the vector of all $1$s in $d$ dimensions. We choose a small value for $\Var(Y_t \st X_t, Z_t)$ because, by Proposition \ref{prop:variance-formula}, that is an irreducible error term that is design-independent. Hence, a small value will allow us to compare design-dependent differences more easily. 

\begin{figure}[t]
    \begin{subfigure}[b]{0.5\textwidth}
        \centering
        \includegraphics[width=\textwidth]{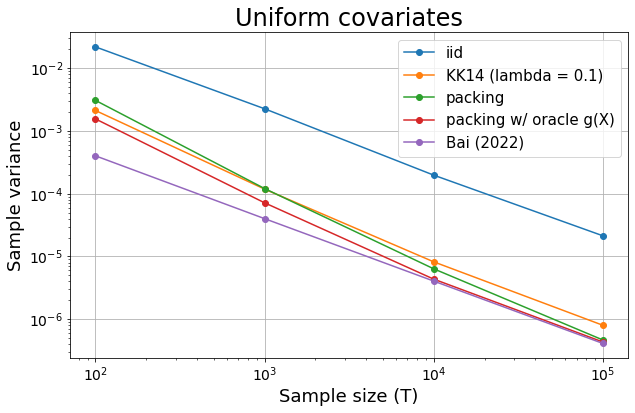}
    \end{subfigure}
    \hfill
    \begin{subfigure}[b]{0.5\textwidth}
        \centering
        \includegraphics[width=\textwidth]{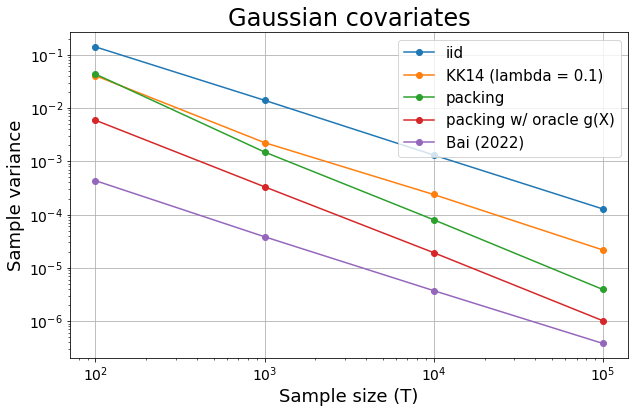}
    \end{subfigure}
\caption{Sample variance of each design (across $1000$ simulations) in both simulation (1) and simulation (2), which use Unif$[0, 1]^2$ and correlated Gaussian covariates in $\bbR^3$, respectively. Both packing-based designs are run with the true sample size $T$ not known a priori, as in Algorithm \ref{alg:packing-design}.}
\label{fig:method-var-comps}
\end{figure}

Figure \ref{fig:method-var-comps} compares the resulting sample variances of $\hat{\tau}_{\IPW}$ across $1000$ simulations, with sample sizes $T \in \{100, 1000, 10000, 100000\}$. In each case, our packing-based method passes the KK14 method and attains a variance nearly identical to that of the \cite{bai2022} method. Unsurprisingly, the IID design performs worst because it is the only design that does not exploit pairing at all. 

In addition, we present an example based on the popular Criteo uplift modeling dataset \citep{criteo}. This dataset is based on online experiments performed by the advertising company Criteo, in which users were randomly selected to see ($Z = 1$) or not see ($Z = 0$) digital advertisements. For the outcome of interest, we choose the binary variable indicating whether the user later visits the advertised site. There are roughly $13$ million rows, though we will not use most of them. Each row also has a vector $X_i \in \bbR^{12}$ of continuous, unlabeled covariates. 

To produce a plausible data-generating distribution, we use $10000$ rows of the dataset to fit logistic regressions for $\bbP(Y = 1 \st X, Z = 1)$ and $\bbP(Y = 0 \st X, Z = 0)$. Given $(X_i, Z_i)$ from the rest of the dataset, we generate $Y_i \sim \text{Bern}(\hat{p}_i)$, with $\hat{p}_i = \hat{\bbP}(Y_i = 1 \st X_i, Z_i))$ being the fitted probability from these logistic regressions. To avoid curse of dimensionality issues, our packing design matches using just the first three principal components of $X$. This assumes that we have strong knowledge of the principal components (for example, via covariate data obtained prior to the experiment). In practice, if this is not the case, one could simply use the empirical principal components up to the current status of the experiment. 

\begin{figure}[t]
    \begin{subfigure}[b]{0.48\textwidth}
        \centering
        \includegraphics[width=\textwidth]{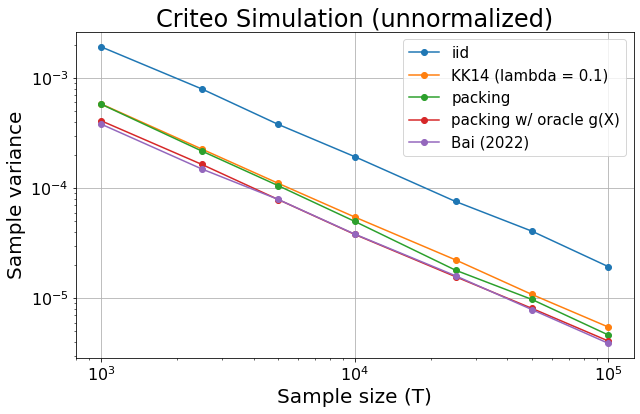}
    \end{subfigure}
    \hfill
    \begin{subfigure}[b]{0.48\textwidth}
        \centering
        \includegraphics[width=\textwidth]{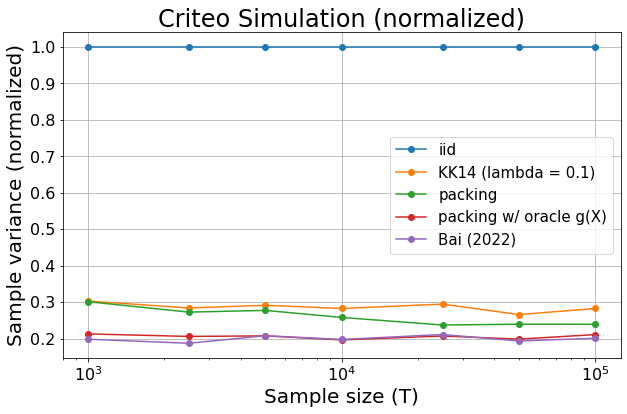}
    \end{subfigure}
\caption{Sample variance of each design for the semi-synthetic example using the Criteo Uplift dataset. Each sample variance is averaged over $2500$ simulations. The left plot shows each sample variance, and the right plot shows them normalized by the sample variance of the IID design at each value of $T$.}
\label{fig:criteo-sim}
\end{figure}

Figure \ref{fig:criteo-sim} shows the resulting sample variances of each design, averaged across 2500 simulations. At each value of $T \in \{1000, 2500, 5000, 10000, 25000, 50000, 100000\}$, we present the sample variances in the left plot and the sample variances normalized by that of the IID design in the right plot. We see that both our method and the KK14 method produce 3-4x improvements over the IID design. For sufficiently large $T$, our packing method also yields a sample variance roughly $10-20\%$ lower than that of the KK14 method.  

Collectively, these simulations highlight the variance improvement of pairing-based designs over IID assignment. Moreover, we have seen here that, for sufficiently large $T$, the packing method produces a meaningful improvement over the method of KK14 by adapting the pairing radius to the current sample size. In addition, we emphasize the desirable property that our method does not have any hyperparameters and is agnostic to the true sample size $T$ and the true distribution of $(X, Y(1), Y(0))$.  

\section*{Acknowledgments} 
The author thanks Art Owen, Adam Kapelner, and Abba Krieger for helpful comments and discussions. 

\newpage 

\bibliographystyle{plainnat}
\bibliography{reservoir}

\newpage 

\appendix 

\section{Proofs} \label{sec:proofs}

\begin{proof}[\bf Proof of \Cref{prop:variance-formula}]
For a given reservoir design, let $n_{\mathcal{M}}(T) = |\mathcal{M}_T|$ and $n_{\mathcal{R}}(T) = |\mathcal{R}_T|$ be the number of matched points and reservoir points, respectively, with $T = n_{\mathcal{M}}(T) + n_{\mathcal{R}}(T)$. Conditional on $\{X_1, X_2, \ldots, X_T\}$, these are deterministic quantities. By decomposing $\hat{\tau}_{\IPW}(T)$ into sums over $\mathcal{M}_T$ and $\mathcal{R}_T$, we obtain
\begin{align*} \Var(\hat{\tau}_{\IPW}(T) \st X) &= \Var\left( \frac{2}{T} \sum_{t = 1}^{T} (2Z_t - 1)Y_t \st X\right) \\ 
&= \frac{4}{T^2} \Var\left(\sum_{t = 1}^{T} (2Z_t - 1)Y_t \st X\right) \\
&= \frac{4}{T^2} \sum_{i \in \mathcal{R}_T} \Var((2Z_i - 1)Y_i \st X_i) + \frac{4}{T^2} \sum_{(i, j) \in \mathcal{M}_T} \Var((2Z_i - 1)Y_i + (2Z_j - 1)Y_j \st X). \end{align*} 
For the first sum, we have
\begin{align*} 
\Var((2Z_i - 1)Y_i \st X) &= \bbE[(2Z_i - 1)^2Y_i^2 \st X] - \bbE[(2Z_i - 1)Y_i \st X]^2 \\ 
&= \bbE[Y_i^2 \st X] - \frac{1}{4} \tau(X_i)^2 \\ 
&= \frac{1}{4} \left( 2\bbE[Y_i(1)^2 + Y_i^2(0) \st X_i] - \tau(X_i)^2\right) \\ 
&= \frac{1}{4} \left(\bbE[Y_i(1)^2 + Y_i^2(0) \st X_i] + \sigma_1^2(X_i) + \sigma_0^2(X_i) + 2 \mu_1(X_i)\mu_0(X_i)\right) \\ 
&= \frac{1}{4} \left(2\sigma_1^2(X_i) + 2\sigma_0^2(X_i) + g(X_i)^2\right).
\end{align*} 
Here, $\tau(X_i) = \bbE[Y_i(1) - Y_i(0) \st X_i]$. For a matched pair, we have that
\begin{align*} 
\Cov((2Z_i - 1)Y_i, (2Z_j - 1)Y_j \st X) &= \bbE[(2Z_i - 1)(2Z_j - 1) Y_i Y_j \st X] - \bbE[(2Z_i - 1)Y_i \st X] \bbE[(2Z_j - 1) Y_j \st X] \\ 
&= -\bbE[Y_i Y_j \st X] - \frac{1}{4} \tau(X_i) \tau(X_j) \\ 
&= -\frac{1}{2} \left(\mu_0(X_i)\mu_1(X_j) + \mu_1(X_i) \mu_0(X_i)\right) - \frac{1}{4} \tau(X_i) \tau(X_j) \\ 
&= -\frac{1}{4} \left(2\mu_0(X_i)\mu_1(X_j) + 2\mu_1(X_i) \mu_0(X_j) +\tau(X_i) \tau(X_j)\right) \\ 
&= -\frac{1}{4} (\mu_1(X_i) + \mu_0(X_i))(\mu_1(X_j) + \mu_0(X_j)) \\ 
&= -\frac{1}{4} g(X_i) g(X_j). 
\end{align*} 
In summary, then, 
$$\Var(\hat{\tau}_{\IPW}(T) \st X) = \frac{1}{T^2} \sum_{i = 1}^{T} \left(2\sigma_1^2(X_i) + 2\sigma_0^2(X_i) + g(X_i)^2\right) - \frac{2}{T^2} \sum_{(i, j) \in \mathcal{M}_T} g(X_i)g(X_j).$$
\end{proof}

\begin{proof}[\bf Proof of \Cref{thm:CLT}]
As in Section \ref{ssec:general-results}, let 
$$\hat{\tau}_{\mathcal{M}}(T) = \frac{1}{n_{\mathcal{M}}(T)} \sum_{(i, j) \in \mathcal{M}_T} (2Z_i - 1)Y_i + (2Z_j - 1)Y_j$$
be the IPW estimator restricted to just the matched set. In addition, consider the design that keeps the same pairs in $\mathcal{M}_T$ but augments them with arbitrary pairings in the reservoir, so that every index is paired (or all but one if $T$ is odd). Let $\mathcal{M}^*_T$ be the set of matched pairs for this hypothetical design, and let $\hat{\tau}_{\IPW}^*$ be the resulting IPW estimator. Since 
\begin{align*} 
\sqrt{T}(\tau - \hat{\tau}_{\IPW}) &= \sqrt{T}(\tau - \hat{\tau}^*_{\IPW}) + \sqrt{T}(\hat{\tau}^*_{\IPW} - \hat{\tau}_{\IPW}),
\end{align*} 
it suffices to show that the desired CLT holds for the first term and that the second term is $o_p(1)$. We begin with the former, for which it suffices to verify that the matchings in $\mathcal{M}^*_T$ satisfy the intra-pair distance condition \ref{eq:intra-distances}. For $r \in \{1, 2\}$, note that
$$\frac{1}{T} \sum_{(i, j) \in \mathcal{M}_T^*} ||X_i - X_j||_2^r = \frac{1}{T} \sum_{(i, j) \in \mathcal{M}_T} ||X_i - X_j||_2^r + \frac{1}{T} \sum_{(i, j) \in \mathcal{M}_T^*\setminus{\mathcal{M}_T}} ||X_i - X_j||_2^r.$$
The first term on the right-hand side is $o_p(1)$ by assumption. For the second term, we take $r = 2$ without loss of generality. Recall that $|\mathcal{M}_T^*\setminus{\mathcal{M}_T}| = n_{\mathcal{R}}(T) = o_p(\sqrt{T})$. Hence, it suffices to show that the maximum $||X_i - X_j||_2^2$ term is $o_p(\sqrt{T})$, for which it suffices by the Cauchy-Schwarz inequality to show that the maximum $||X_i||_2^2$ term is $o_p(\sqrt{T})$. Recall that $\bbE[||X_i||_2^4] < \infty$ by Assumption \ref{assumptions:XY}. Then, by a standard result in probability (e.g., Lemma 6 of \cite{morrison-owen25}), $\underset{i \leq T}{\max} \text{ } ||X_i||_2^2$ is $o_p(\sqrt{T})$. This establishes condition \ref{eq:intra-distances}, so 
$$\sqrt{T}(\tau - \hat{\tau}^*_{\IPW}) \overset{d}{\to} N(0, \sigma^2_{\red}).$$
For the second term, we let $(Z_i, Y_i)$ be the treatment and outcome from the original reservoir design, and we let $(\tilde{Z}_i, \tilde{Y}_i)$ be the treatment and outcome from the design that pairs units in the reservoir. Then
$$\sqrt{T}(\hat{\tau}^*_{\IPW} - \hat{\tau}_{\IPW}) = \frac{1}{\sqrt{T}} \sum_{(i, j) \in \mathcal{M}_T^*\setminus{\mathcal{M}_T}} (2\tilde{Z}_i - 1)\tilde{Y}_i + (2\tilde{Z}_j - 1)\tilde{Y}_j - \frac{1}{\sqrt{T}} \sum_{i \in \mathcal{R}_T} (2Z_i - 1)Y_i,$$
where both sums on the right-hand side include the same indices (those in the original reservoir). There are $o_p(\sqrt{T/\log(T)})$ terms in each sum by the assumption on the reservoir size, and each summand is $O_p(\sqrt{\log(T)})$ by the sub-Gaussian assumption on the potential outcomes. Hence, each sum is $o_p(1)$, which concludes the proof.  
\end{proof}

\begin{proof}[\bf Proof of \Cref{thm:CLT-packing}]  
Let $A_T = \{X_t \in [-\log(T), \log(T)]^d \text{ for all } t \leq T\}$. Because $X_i$ are IID and sub-Gaussian, the maximum coordinate of any one of them in the first $T$ is $O_p(\sqrt{\log(T)})$. Hence, $\underset{T \to \infty}{\lim} \text{ } \bbP(A_T) = 1$. 

Suppose first that the procedure of Algorithm \ref{alg:packing-design} is run with $T$ known and $\lambda_T = T^{-1/(2 + \delta)d}$. Conditional on the event $A_T$, the size of the reservoir is $O\left(\log(T)^d/\lambda_T^d\right)$ because the $\lambda_T$-packing number of $[-a, a]^d$ is $O(a^d/\epsilon^d)$. Hence, the reservoir size is 
$$O_p\left(\log(T)^d/\lambda_T^d\right) = O_p\left(\log(T)^d T^{1/(2 + \delta)}\right) = o_p\left(\sqrt{T/\log(T)}\right).$$
Now suppose that $T$ is not known a priori, so the cutoff radius at index $t$ is $t^{-1/(2 + \delta)d}$, as described in Algorithm \ref{alg:packing-design}. The covariates of the units in the final reservoir will still form a $\lambda_T$-packing. To see this, suppose that units with index $t_1 < t_2$ are in the final reservoir. If $||X_{t_1} - X_{t_2}||_2 < \lambda_T$, then $||X_{t_1} - X_{t_2}||_2 < \lambda_{t_2}$ as well, and so they would have been paired when unit $t_2$ entered. Hence, we have the same bound on the reservoir size. 

Because the intra-pair distances converge to zero by construction as well, we can apply Theorem \ref{thm:CLT} to conclude.
\end{proof}

\end{document}